# 3D printing of nanoceramics: Present status and future perspectives

**Mainak Saha[1,2]**

[1]Department of Metallurgical and Materials Engineering, Indian Institute of Technology, Madras-600036, India

[2]Department of Metallurgical and Materials Engineering, National Institute of Technology, Durgapur-713209, India

Corresponding author: Mainak Saha

Email-address: mainaksaha1995@gmail.com

**Abstract**

In materials research involving additive manufacturing (AM)-based techniques for fabrication of a wide variety of materials, the latest trend at present is to focus largely on 3D printing (3DP) of nanoceramics, which at present is highly challenging, from both fundamental and industrial viewpoints inspite of the tremendous versatility offered by these techniques in terms of addressing design complexities [1]–[6]. The two main reasons for the same are: (i) low density and (ii) poor mechanical properties of nanoceramic parts fabricated using 3DP techniques [7]–[9]. The fundamental reason behind the two aforementioned features of 3DP-fabricated nanoceramic parts is the huge extent of microstructural inhomogeneity arising primarily due to variation in cooling rates during 'point by point', 'line by line' or 'layer by layer' deposition methodology followed in 3DP techniques [10]–[16], leading to a number of defects in the microstructure [17], [18]. Moreover, the industrial application of nanoceramic parts manufactured using 3DP techniques, is rather limited, primarily owing to the high manufacturing cost associated with these nanoceramic parts. Although, in the last ten years, there has been a considerable volume of work on 3DP-based techniques for manufacturing ceramic parts with enhanced densities and improved mechanical properties, however, there is limited understanding on the correlation of microstructure of 3DP-fabriated nanoceramic components with the mechanical properties. On the other hand, in the recent decade, the 'correlative' methodology of characterising microstructures from micro to nanoscale, involving a number of different structural and chemical characterisation techniques, for the study of a number of defects ranging from the equilibrium point (or 0-D) to non-equilibrium volume (or 3-D) defects, has been hugely employed in a number of metallic materials [19]–[21]. This has completely revolutionised the understanding of structure-property correlation and microstructural defects in these materials and paved a whole new dimension towards a systematic correlation of structure (ranging from bulk to nano-scale) to a wide range range of properties in these materials. However, in the context of 3DP-fabricated nanoceramic parts, at present, there is hardly report on understanding structure-property correlation using the aforementioned methodology. The present review is aimed to review some of the most commonly used 3DP techniques for the fabrication of nanoceramics and provide an overview of the future perspectives, associated with the necessity towards developing a systematic structure-property correlation through 'correlative' characterisation methodology in these materials.

# 1. Introduction

Ceramics, with dimensions ranging from a few nanometers to a few microns have been reported to undergo brittle to ductile transition at room temperature rendering them as ideal candidates for a number of engineering applications [22]. Moreover, properties such as high stiffness coupled with high resistance to high temperatures and chemical attack, and low density, would additionally render ceramics as ideal candidates for a number of structural engineering applications, which ranges from automotive to aerospace systems [23]. However, the primary limitation of ceramics is their brittle nature which presently, restricts their application in structural components [22]. In addition, the processing of ceramics is challenging as compared to that of metallic materials [23], [24]. The brittleness of ceramics is derived from microstructural porosity, cracks, and a number of different inhomogeneities induced during processing of these materials.

Nanoceramics have been defined as a type of nanoparticles comprising of ceramics, which are inorganic, heat-resistant, and non-metallic in nature [6]. On a macroscale, ceramics are brittle and rigid [24], [25]. Recently, pyrolysis of additively manufactured (AM) complex-shaped parts into low-flaw-population amorphous silicon oxycarbide (SiOC) using UV-curable preceramic resins has been reported [22]–[24]. However, the present investigations on mechanical response are limited to specimens with characteristic length scales of the order of a few millimeters, with brittle fracture as the primary mode of failure [24]. At present, application of ductile nanoceramics has been primarily limited to focused ion beam (FIB) based milling of thin films [22]. In terms of AM technique with the highest possible three-dimensional (3D) resolution, two-photon polymerization direct laser writing (TPP-DLW) technique with pre-ceramic resins have been reported to provide a pathway towards fabrication of ductile nanoceramics [24]. Till date, the only TPP-DLW-derived ceramic is pyrolytic glassy carbon (C) [22], [24]. Pyrolysis of properly designed polymeric microstructures has been reported to create extremely strong glassy C nanostructures [22], [24]. However, brittleness with a number of scattered properties related to size, geometry, and fabrication techniques, and shrinkage of nearly up to 90% upon pyrolysis, have been reported. In addition, scattered plastic response has been observed in pillar compression experiments [22]. However, the effect has been reported to disappear with increase in pillar diameter to micrometer scale leading to a massive degradation in mechanical properties. At present, application of high-strength TPP-DLW-derived glassy C appears to be limited to materials used for architectural purposes [22].

In the context of regenerative medicine and tissue engineering, nano-scale applications have been extensively studied due to the nano-sized nature of interactions between cells and the extracellular matrix (ECM) of tissues [6]. In the field of biomedicine, nanoceramics have been reported to provide enormous potential for nanomedical devices (for example, sensors for diagnosis and monitoring of diseases) [25], [26]. At present, the limitations of autografts and allografts have led to extensive research on synthetic grafts [27]. Moreover, applications based on nanotechnology, have been extensively studied due to the nanosized nature of interactions between cells and tissues [6].

In the context of electromechanical devices such as actuators, sensors and transducers generating electric charge from an applied mechanical impulse, and vice versa, piezoelectric ceramics play an essential role [28], [29]. The most commonly used piezoelectric materials are based on lead zirconate titanate ($PbTiO_3$ or commonly abbreviated as PZT), owing to their high piezoelectric constants [30]–[32]. However, Pb owing to various environmental hazards arising out of its toxicity, is an essential constituent in these materials [28], [32]. The recent advancement in this field is the development of Pb-free piezoelectric materials such as Potassium-Sodium Niobate (KNN) and Barium Titanate ($BaTiO_3$) as alternatives to PZT ceramics [28]. $BaTiO_3$ ceramics, in particular, have been reported to be potential candidates for applications ranging from chip capacitors to medical implants [28], [33]. A number of conventional fabrication techniques have been developed to manufacture piezoelectric materials with different geometries, which commonly include laser or ultrasonic cutting, etching and dicing, injection molding and jet machining [28]. Although these techniques have been reported to manufacture components with complex geometries, they are highly expensive and lack dimensional accuracy [28]. Moreover, mechanical stresses in case of traditional processing methods may lead to a number of phenomenon such as depolarization of the near-surface area, leading to a significant degradation of piezoelectric properties [28]. To address the limitations of conventional fabrication techniques, AM-based techniques have led to an effective pathway towards the development of low-cost fabrication techniques for manufacturing complex-shaped piezoelectric components with high dimensional accuracy [28].

In recent times, the novel "correlative microscopy" methodology [34]–[37] involving a number of structural and chemical characterisation techniques from the same region in a microstructure, has been extensively employed as a tool to correlate crystal structure with a number of properties in metallic materials. One of the primary reasons as to why the aforementioned methodology has not been used for nanoceramics may be attributed to limited understanding of microstructure in nanoceramics. The reason for the same may further be attributed to limited research from the viewpoint of materials science. The present chapter limits its premises in discussing the state-of-the-art on research in the field of nanoceramics and finally, some of the major challenges, requiring systematic research, from both industrial and fundamental viewpoints, which, in future may aid in creating a new paradigm in the field of research on nanoceramics.

## 2. Two-photon polymerization direct laser writing (TPP-DLW)

Pyrolysis at $1,000°C$ has been reported to yield undistorted ceramic SiOC structures with feature sizes of approximately ~200 nm **(Figure 1(b))** [22]. The amount of linear shrinkage upon pyrolysis has been reported to be nearly 30%, which is in line with larger-scale additively manufactured SiOC [22], [38], [39]. Octet nanolattices, woodpile photonic crystals and monolithic micropillars pre and post pyrolysis, have been shown in **Figure 1(a-c)**.

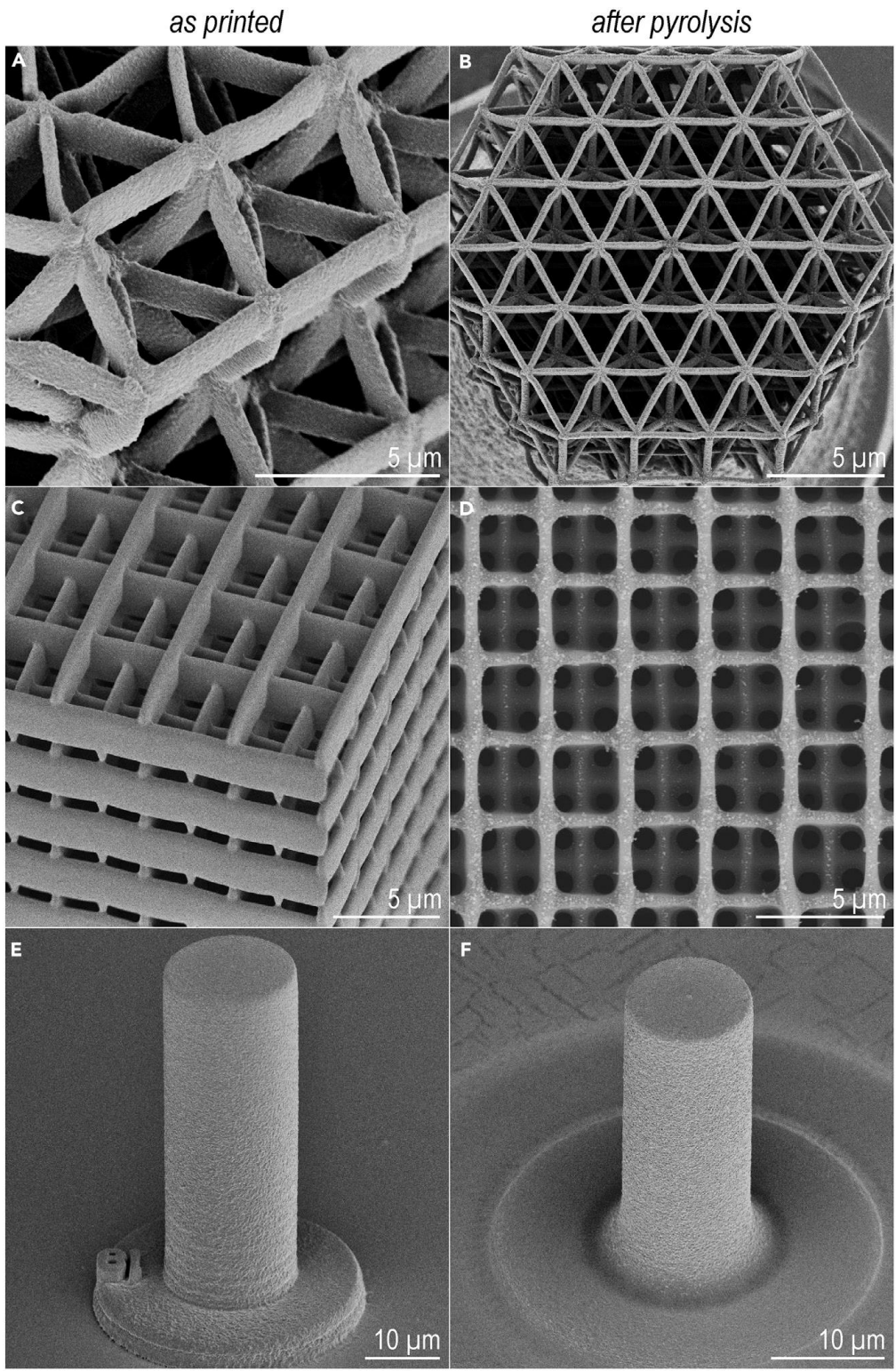

**Fig 1.** AM of SiOC Nanoceramics via TPP-DLW and Pyrolysis: **(a and b)** 3DP polymeric and pyrolyzed SiOC octet nanolattice, **(c and d)** woodpile photonic crystal, and **(e and f)** micropillar [22].

**Figures 2(c–l)** represent different-sized specimens in pre-deformed and post-failed conditions. Uniaxial compression of TPP-DLW-derived SiOC (TPP-SiOC) micropillars (with diameters ranging from 1 to 20 mm) have been reported to consistently show ductile deformation behaviour with failure strains up to 25% and ultra-high compressive strengths, very close to theoretical limit (~E/10) (**Figure 2(a)**) [22]. Moreover, SiOC micropillars synthesized using TPP-DLW technique have been reported to possess extremely high yield strength (~5-7 GPa) and elastic modulii (~67 GPa) with ductility of nearly 9-15% [22]. An interesting deformation mechanism without the presence of shear bands has also been observed [2], [22]. In addition, it has also been reported that during deformation, there is nucleation of longitudinal cracks leading to vertical splitting. With decreasing specimen size, the top surfaces of the pillars have been reported to become conical-shaped owing to a number of constraints during fabrication process (**Figs. 2(c and d)**) [22]. Additionally, for specimens with diameters less than 2 mm, the above defect (geometric in nature) leads to early nucleation of crack and consequent reduced strength and Elastic modulii [22].

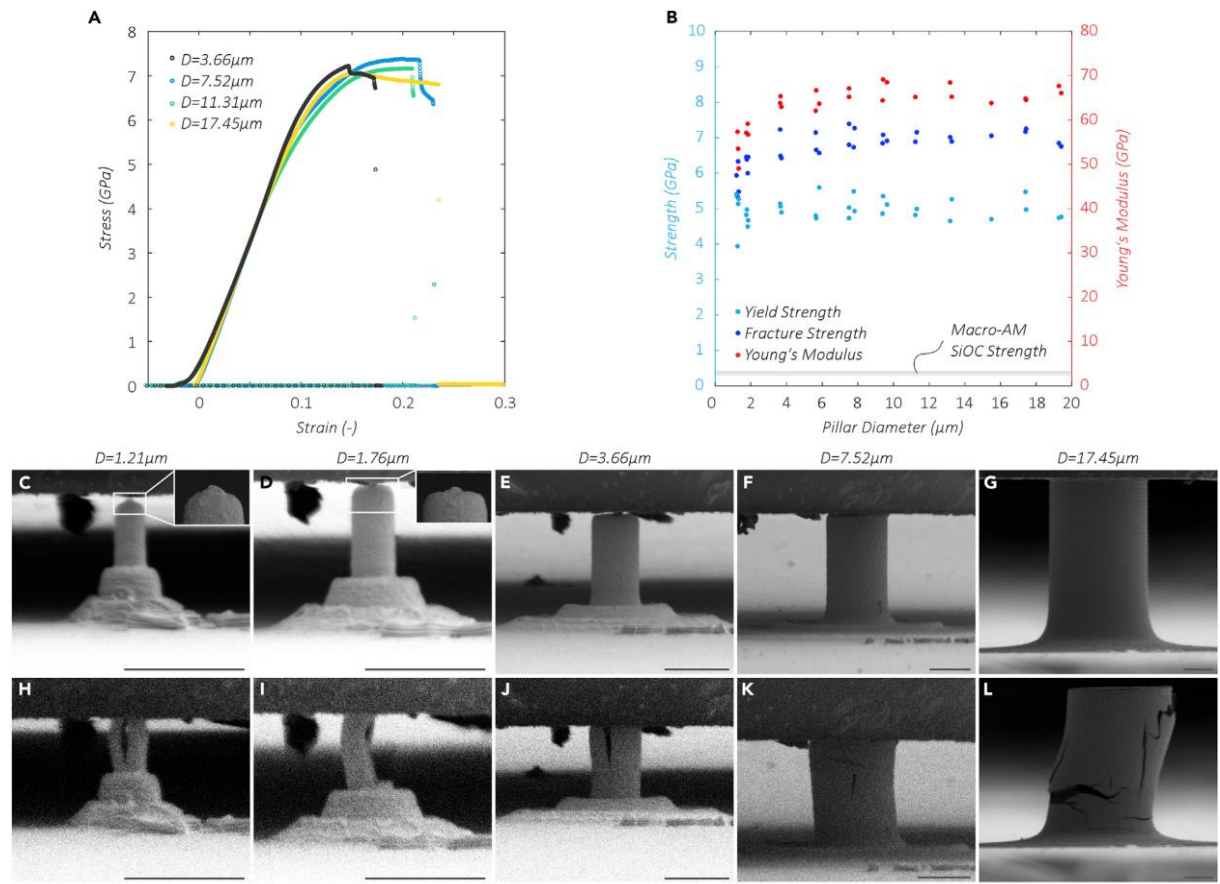

**Figure 2** Compression of SiOC micropillars: **(a)** Stress-strain curves showing ductile deformation behaviour as a function of pillar diameter (D). Variation of **(b-l)** Uniaxial yield

strength ($\sigma_y$), fracture strength ($\sigma_f$), and Elastic modulii (E) with D: **(b)** SEM images of different-sized pillars in **(c–g)** pre-deformed and **(h–l)** post-failed conditions. Scale bar of 5 mm has been used from parts **(c-l)** [22].

Transmission electron microscopy (TEM) based investigations of the TPP-SiOC reveals a completely amorphous pore-free microstructure (**Figure 3**). Using Energy-dispersive X-ray spectroscopy (EDS), a uniform distribution of Si, C, O, and S (**Figure 3(d)**) has been reported which is comparable with that of previously reported polymer-derived SiOC [22], [40]. Moreover, SiOC octet nanolattices have been reported to be the toughest metamaterials, to date [40]. Here, it becomes essential to mention that EDS technique working on the principle of characteristic X-Ray energies from different elements cannot be used for determination of accurate concentration of light elements with lowatomic numbers (Z) such as B, C, N, O etc. as reported in an extensive review on EDS technique by Cerqueira et al. [41]

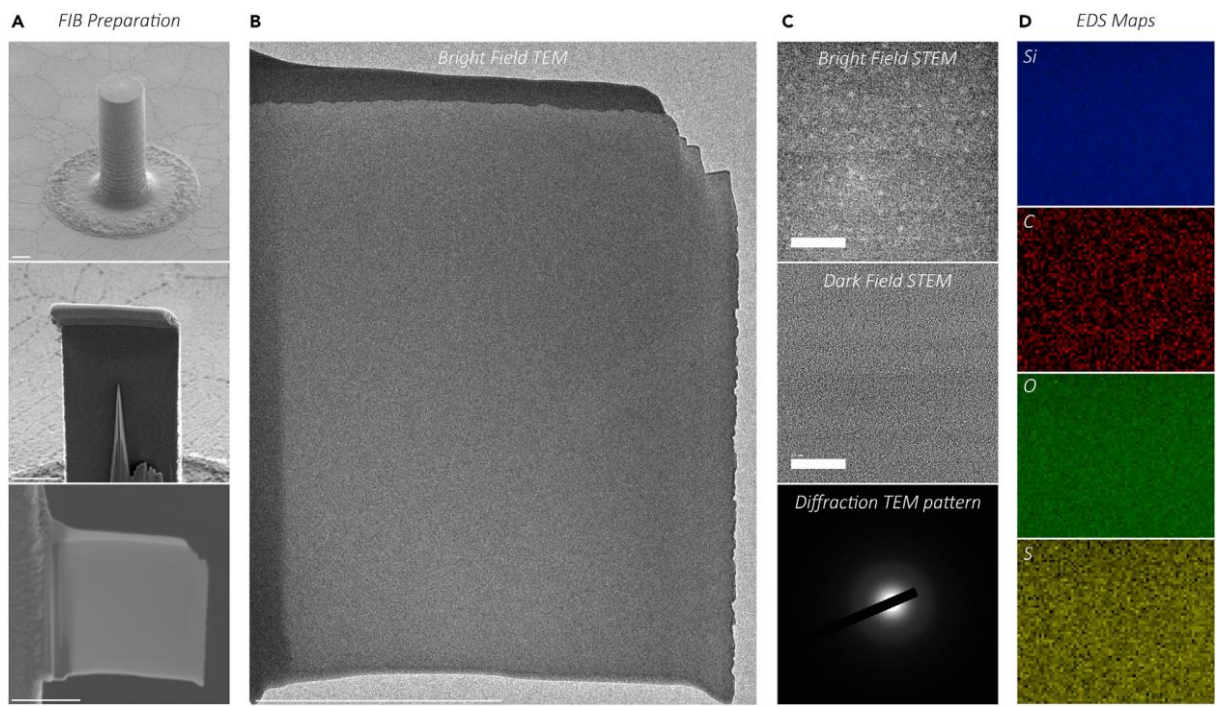

**Figure 3** Microstructural characterization using TEM: (**a and b**) FIB-based extraction of sample from the centre plane of a micropillar using bright-field TEM images in **(a)** and **(b)**. Scale bars of 5 mm have been used in **(a)** and **(b)**. (**c**) Bright- and dark-field STEM images and selected area diffraction patterns (SADP) confirming amorphous pore-free structure. (**d**)

Energy-dispersive X-ray spectroscopy (EDS) maps showing uniform distribution of Si, C, O, and S. Scale bars of 10 nm have been used in **(c)** and **(d)** [22].

There are two primary mechanisms governing plasticity (during compressive deformation) in amorphous materials: **(i)** volume-conserving shear flow (occurring through localized bond-switching events) [42], and **(ii)** volumetric strain (through irreversible densification which is dominant in the case of very open structures) [43]. The structure of amorphous SiOC may be visualised as a random network of tetrahedral $SiO_4$ units, similar to that in Silica ($SiO_2$) glass, with the replacement of some O by C atoms [2]. Moreover, volume-conserving shear flow has been reported to be hugely dominant during compression of TPP-SiOC, similar to that observed in $SiO_2$ glass, mainly due to a high level of similarity in terms of crystal structures of both, as discussed in the previous statement [43].

Neither shear flow nor densification have been reported to be intrinsically dependent on size, although both require extremely high activation stresses (~5GPa, as reported by Bauer et al. [22]), which are practically unapproachable owing to a low fracture toughness [44]. Fracture strength, on the other hand, is largely dependent on the processing-based microstructural defect population and has been widely reported to increase with decreasing dimensions [43], [45], [46]. The size-independent low-defect population of TPP-derived SiOC, leads to a high overall toughness i.e. high ductility and strength simultaneously [22]. Short diffusion paths during 3DP may lead to a lower inner-material flaw population than in stereolithography technique involving polymerisation of a large volume of material [43]. In contrast to TPP-fabricated ceramics (e.g. pyrolytic glassy C), TPP-SiOC has been reported to be largely insensitive towards surface-to-volume effects. There is a large amount of decomposition involved in the decomposition of polymers to glassy C, leading to extreme shrinkages [44]. Besides, with increasing dimensions, there is an increase in average diffusion path length for molecules to move out of the material leading to void formation and consequent degradation of mechanical properties [43]. On the contrary, the transformation of preceramic polymer to SiOC, has been reported to lead to low shrinkage (~30%) [45]. In addition, the fully amorphous microstructure of TPP-SiOC results in almost uniform mechanical properties. On the other hand, surface-induced graphitization in pyrolytic glassy C has been reported to produce a considerable size effect leading to a scatter in mechanical properties [46].

3. **Crack propagation resistance of 3DP nanoceramics**

3DP fabricated nanoceramic metamaterials have recently emerged as a new class of lightweight materials with exceptionally high strength and stiffness [47], [48]. However, the application of these materials is presently limited owing to limited information on the mechanical properties of these materials [49], [50]. In addition, the mechanical behaviour of AM-based pyrolytic C is complicated [47]. For glass ceramics, humidity has a deleterious influence on the fracture strength, primarily leading to simultaneous action of multiple events, including chemical reactions between the Si–O–Si bond and water at the crack tip [47], [51]. Using the double-cleavage-drilled-compression test in an environment with relative humidity (RH) ranging from 50 to 56%, it has been shown that crack propagation velocity varies as a function of fictive temperatures of the glass, thereby confirming that crack growth occurs due to reaction of water (from the environment) with the glass [52], [53]. Moreover, at present, only few studies have been based on understanding the influence of humidity on pyrolytic glassy C which, besides, is highly insensitive to fatigue in water. Glassy C with high elastic modulus have been reported to be crack resistant [54]–[57].

Rossi et al. [47] have recently characterized the fracture toughness of TPP-DLW-fabricated pyrolytic C via nanoindentation micro-pillar splitting technique for understanding the influence of humidity on material's surface flaw distribution. This technique has been reported to involve the indentation of micro-pillars uptil fracture [58]. Besides, there is no requirement of any post-test measurement of crack length [47]. Fracture toughness ($K_c$) as a function of failure (splitting) load $P_c$ and radius of pillar radius R is determined by the following relationship: [2], [58], [59]:

$$K_c = \gamma \frac{P_c}{R^{\frac{3}{2}}}. \qquad (1)$$

where, ɤ is the calibration coefficient and has been reported to be a dimensionless constant depending on a number of factors such as hardness to elastic modulus ratio, indenter geometry and Poisson's ratio of the material [58]. ɤ has been reported to be independent of pillar size when the average grain size is much smaller than pillar diameter [2]. **Figure 4** shows the Scanning Electron Microscope (SEM) images of pyrolytic C micro-pillars with and without the presence of flaws on the surface.

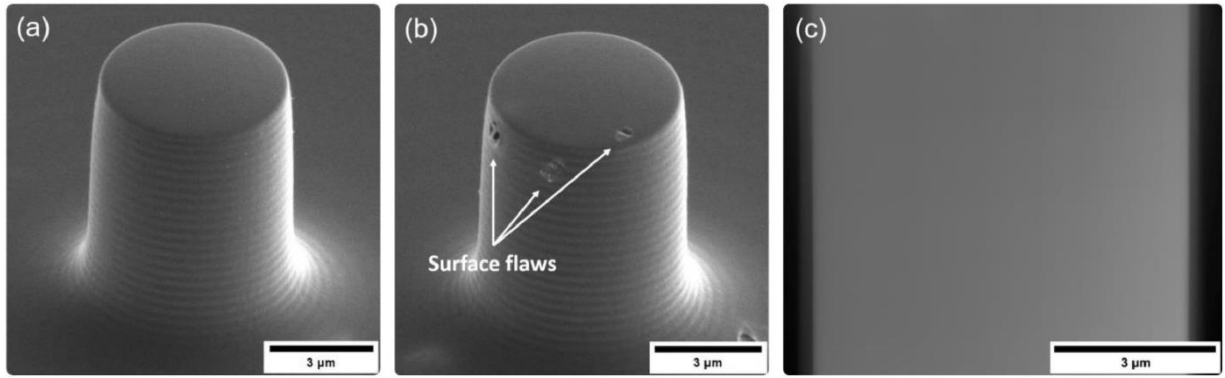

**Figure 4** SEM images of pyrolytic C micro-pillars in **(a)** absence and **(b)** presence of observable flaws on the surface, respectively. STEM image of an electron transparent lamella extracted from the vertical cross-section of a pillar using FIB based liftout technique [47].

**Figure 5** illustrates the nanoindentation response, reporting the variation of elastic modulus (E) and hardness (H) as functions of indentation depth at two different RH levels: RH <5% and RH >60%, for pyrolytic C micro-pillars (of diameter ~ 11.8 µm). E was observed to be independent of RH.

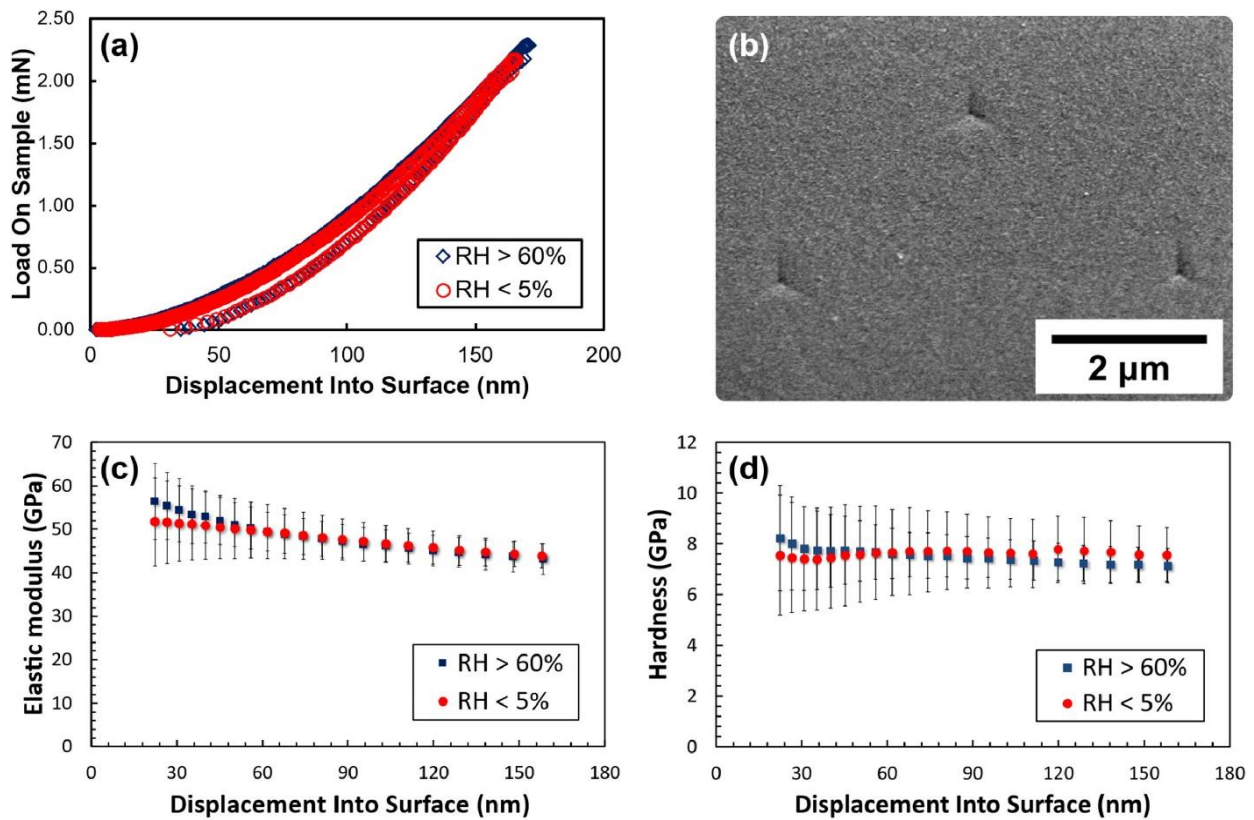

**Figure 5** Nanoindentation-based (a) load vs depth curves and (b) Berkovich indents in the microstructure (c) Variation of elastic modulus (E) as function of depth and (d) Variation of hardness as function of depth profiles at relative humidity (RH) values of <5% and >60% [47].

**Figure 6** illustrates the splitting experiments with pyrolytic C micro-pillars [47]. An arrangement of the splitting procedure with a specimen with surface flaws is given in **6(a)** [47]. Indentation marks in a pre-fractured condition (at a load of 50 mN), showed the presence of surface concentric ring-shaped crack (with $45°$ orientation) on the specimen surface, as shown in **6(b)** [47]. FIB analysis of indents revealed the presence of median cracks (**6(c)**) [47]. Based on the study, median cracks were reported to be the main driving force for failure. **6(d and e)** represent load vs depth plots for pristine structures and specimens with surface defects measured at RH values of <5% and >60%, respectively [47]. In all cases, 'pop-in' events representing displacement burst were observed suggesting the occurrence of failure through unstable crack propagation [47].

The influence of humidity on the fracture toughness of TPP-DLW-fabricated pyrolytic C may be explained by an interplay between RH-level and the distribution of flaws in the material. This is largely scale-dependent. It has been reported that pyrolytic C is hydrophilic for water contact angle in the range of $50-70°$, suggesting that at high RH levels, the presence of surface flaws aids diffusion of water into the micro-pillar specimens (**Figure 7**) [60]–[62]. Besides, chemical reactions between water and glassy C have been reported to induce local weakening of the material during pillar splitting experiments [47]. Although decreasing flaw size has been reported to induce toughening, however, the influence of chemical modifications (on the specimen surface) increases with decrease in specimen size [47]. Moreover, capillary effects from water have been attributed to lead to a change in the distribution of stress fields around surface flaws [47].

## 4. Nanoceramics for biomedical purposes

Nanoceramics for tissue engineering/regenerative medicine, have been classified as bioactive, bioresorbable, or bioinert [6].

Bioactive glass nanoceramics (nBG): Upon degradation, they release ions for promoting osteogenesis and angiogenesis [6]. Besides, these materials convert to a biologically active carbonated apatite which firmly binds itself to bone [6]. Yuan et al. [63] have attributed osteoinductivity (in 45S5 bioactive glass) to the formation of dissolution products, stimulating osteoprogenitor cells at the genetic level [63]. Bioactive glasses find a wide range of biomedical applications, ranging from bone repair and regeneration, to skin repair [6], [27], [63].

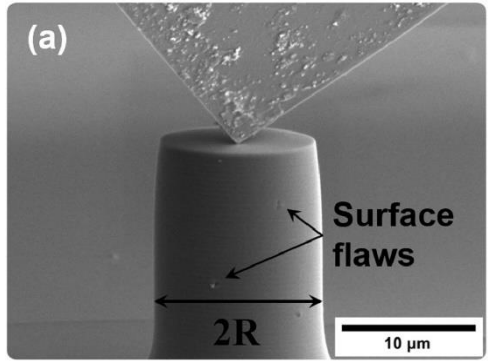
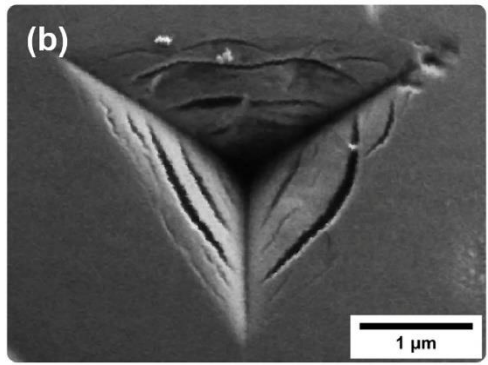
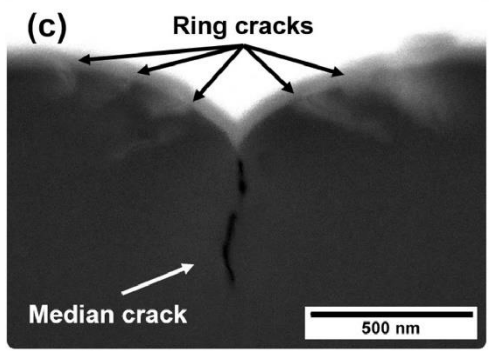
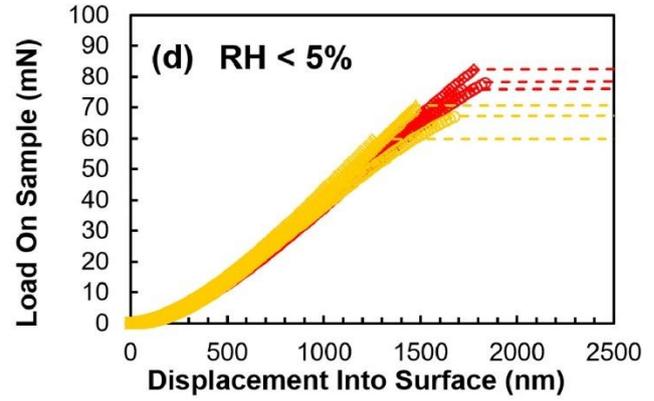
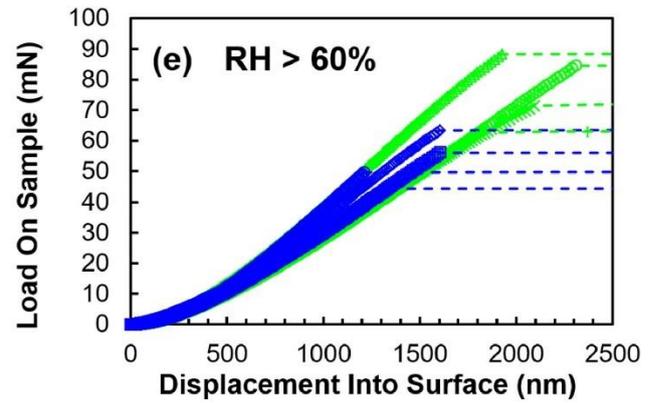
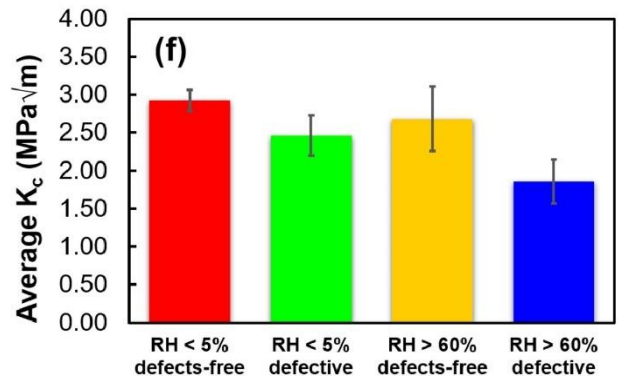

**Figure 6** (**a**) Arrangement of pillar splitting testing procedure on a pyrolytic C micro-pillar with surface flaws and (**b**) observed morphology of surface cracks before the splitting. (**c**) Post-mortem SEM image of Berkovich indent (along the depth). Nanoindentation-based load vs displacement plots at RH levels of (**d**) <5% and (**e**) >60%. Part (**e**) shows a comparison of mean fracture toughness at two RH levels ranging from pristine to defected micro-pillars. Colours in part (**f**) correspond to colours in parts (**d**) and (**e**) [47].

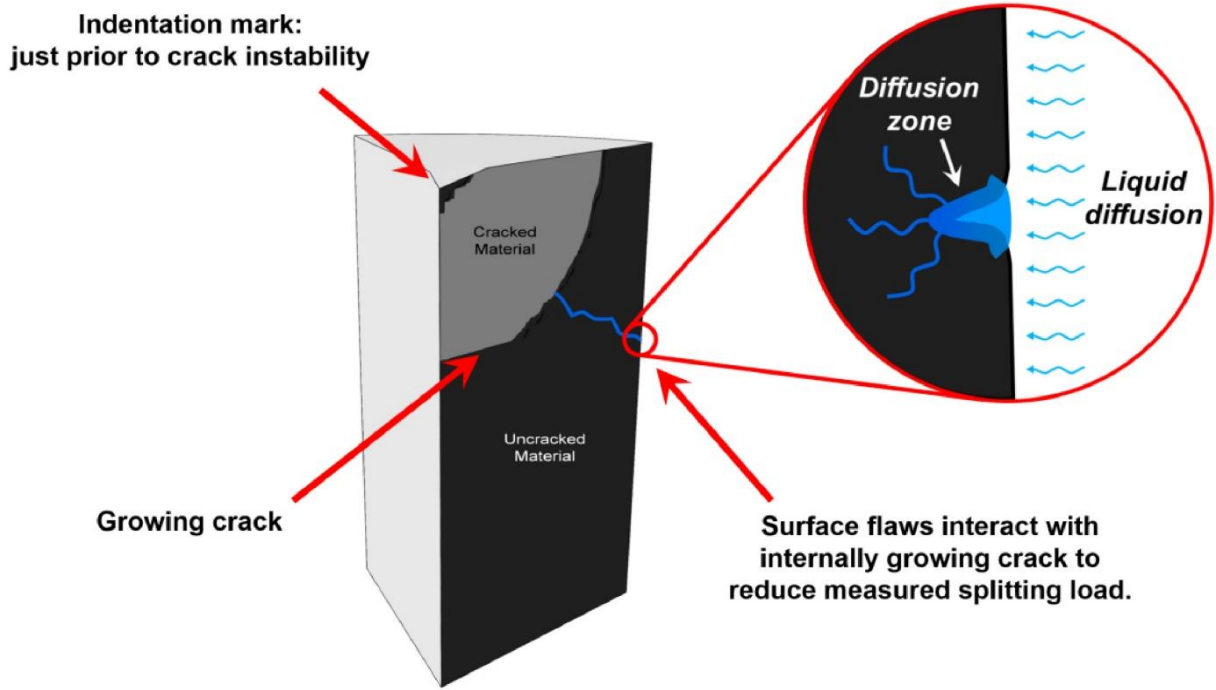

**Figure 7** Schematic representing the interaction between main crack (induced during nanoindentation) and surface flaws leading to an increase in stress intensity factor and, consequent, reduction of critical splitting load [47].

Bioresorbable nanoceramics: These are calcium-phosphate (CaP) materials including hydroxyapatite (HA), tricalcium phosphate etc. [6]. These materials have been reported to possess enhanced biocompatibility, bioactivity, osteoconductivity, and bioresorbability owing to their chemical similarity with the tissues present in the human body [64]. Moreover, these materials form a bioactive apatite layer on the surface of bone tissues, thereby forming a direct bond with these tissues and enhancing their osteointegration [27], [64]. Some of the CaP materials have even been reported to be highly osteoinductive [27]. Solubilization and resorption of CaP materials are dependent on four factors viz. pH of solution, composition, temperature and microstructural features such as topography of material, particle size, and pore size [65]. During hydration, exposure of nano CaP materials to biological fluids have been reported to lead to release of ions like $PO_4^{3-}$, $Ca^{2+}$, and $HPO_4^{2-}$, regulating the functioning of osteogenic cells . Hence, these materials are reinforced with polymers to form composites [27].

Bioinert nanoceramics: These are Ti, $Al_2O_3$, and $ZrO_2$-based materials and are characterized by high bioinertness, fracture toughness, and strength [6]. For instance, Ti and its alloys manly find applications in reconstruction of bone tissues owing to their excellent corrosion resistance [6].

Bioceramics, in particular, may be classified as oxides or nonoxides. $Al_2O_3$, $ZrO_2$, and $TiO_2$ may be placed under the oxide category whereas SiC and $Si_3N_4$ fall under non-oxide category (**Figure 8**). They are mainly composed of phosphates, silicates, or carbonates. Due to their physicochemical properties, they find extensive applications as biomaterials for tissue engineering applications.

A variety of techniques have been reported for the fabrication of bioactive nanoceramics. These are primarily classified as top-down and bottom-up processes[27]. A top-down (TD) approach involves dissociation of bulk material to the required structural form, whereas a bottom-up (BU) approach involves adding up material from further smaller pieces to obtain the required structure [6]. A number of TD and BU techniques have been reported [27]. For instance, patterning, and comminution techniques involving TD approaches. Nanolithography, nanoimprint, and nanoprinting are common patterning techniques [65]. Physical and chemical vapor deposition (PVD and CVD respectively) and atomic vapor deposition are some common examples of additive techniques [27]. Dry and wet etching are common subtractive techniques whereas grinding and milling may be placed under comminution techniques [6]. While TD methods are highly cost-effective, BU techniques are preferred over TD methods for a higher fabrication quality and enhanced structural homogeneity [6], [66]. In the context of fabrication of bioactive nanoceramics, phase separation is a very commonly reported technique involving BU approach [6], [67].

Hong et al. [68] have reported the advantages and disadvantages of various fabrication techniques for CaP based nanoceramic powders and coatings combined with their biological characteristics. Wet-chemical synthesis and sol-gel technique have been reported to be the most commonly used BU techniques for the fabrication of various ceramic nanoparticles including CaP, iron oxides, $TiO_2$, etc. [6]. Sol-gel technique involves hydrolysis and polycondensation reactions [27]. To be more specific, sol-gel technique involves preparation of a mixture of precursors, which undergo subsequent transformation into the final product (through drying, gelation, and curing). The main advantages of this technique include lower processing temperatures, high purity of the fabricated products, and the ability to fabricate multicomponent materials in a number of different forms [6].

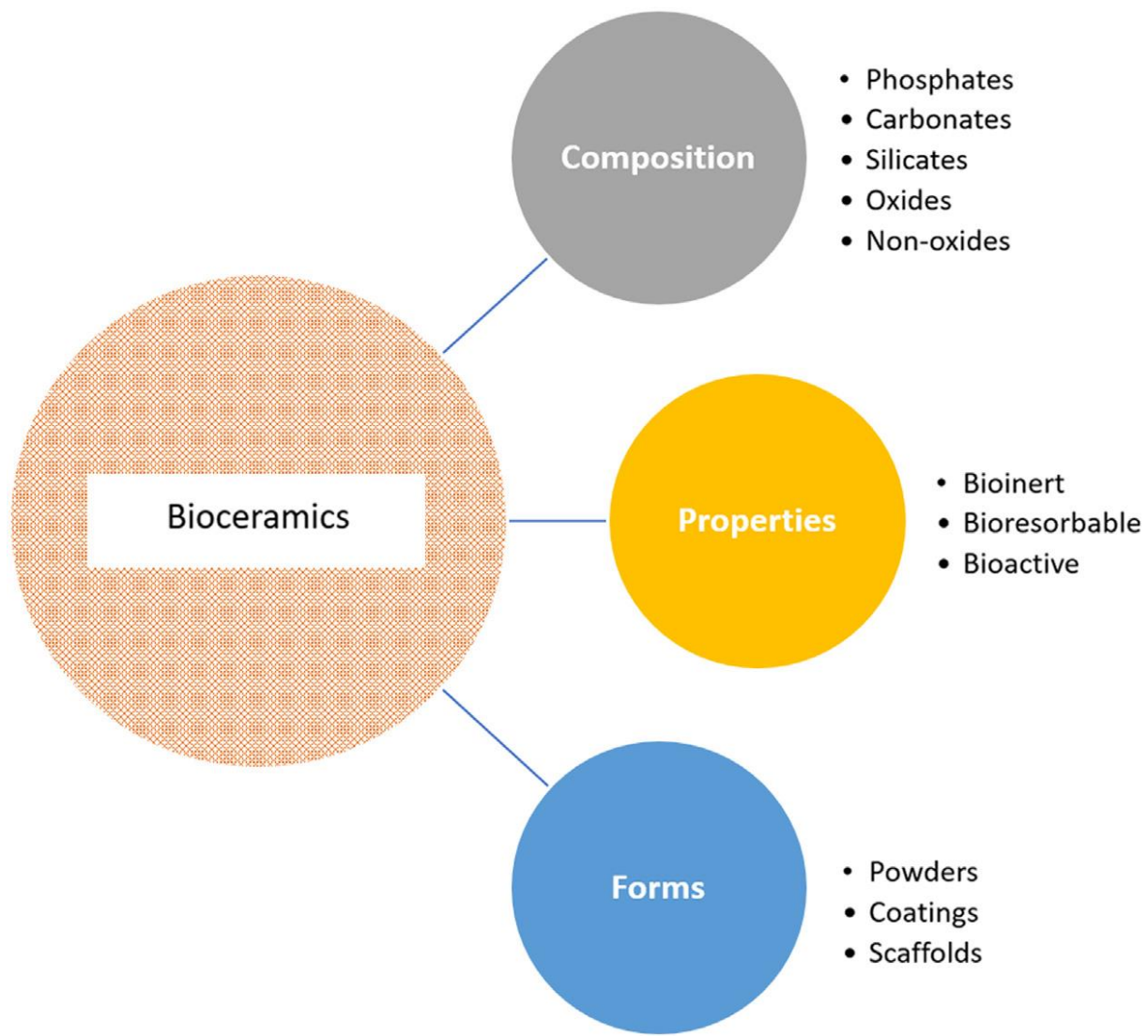

**Figure 8** Schematic representing classification of bioceramics [6].

Various fabrication techniques have been reported to result in different shapes and sizes of nanoceramics with a number of different surface areas [27], [65]. The properties and potential applications of CaP are highly dependent on microstructural features such as grain size, shape, and distribution as well as on the way of deposition which may either vary in the form of coating or powder [65], [67]. The use of nano-sized crystals of HA (width ranging from 5 to 20 nm) has been reported in the bone tissues. Synthetic nano-hydroxyapatite (nHA) is extensively used in the form of powders, granules, and porous blocks in the orthopedic sector.

Another commonly studied bioactive nanoceramic is Bioglass 45S5 [67]. It is a multicomponent oxide glass comprising of $SiO_2$, $Na_2O$, $CaO$ and $P_2O_5$. Most importantly, it has low SiO$_2$ content with high Na$_2$O and CaO contents with a high CaO/P$_2$O$_5$ ratio [27], [67]. In addition to the silicate bioglasses, the other common bioactive glasses are phosphate-based and borate-based. The two main methods for fabricating bioactive glasses are conventional melt-quenching method or the sol-gel method (**Figure 9**) [6]. During melt quenching technique, different quantities of raw materials such as SiO$_2$, Na$_2$CO$_3$, CaCO$_3$, and Ca$_2$P$_2$O$_7$ are mixed initially, followed by subsequent melting at temperature ranging from 1300°C to 1450°C and annealing between 450°C-550°C [6]. On the other hand, the first step in sol-gel synthesis of bioactive glasses involves mixing of alkoxide or organometallic precursor followed by their hydrolysis, resulting in the formation of silanol groups which interact with each other to form SiO$_2$ network through polycondensation reactions [6]. This is followed by gelatination [6], [27]. With passage of time, there is a 3D network developed between a large number of particles leading to the formation of a highly viscous liquid (gel) [6]. This is followed by the ageing of the gel through a number of polycondensation and reprecipitation reactions [6]. This is followed by the stabilization and sintering of the aged gel [6].

The selection criteria for the most suitable technique for the fabrication of Bioglass depends on a number of factors with the overall objective is to obtain a composition with controlled bioactive behaviour [27]. The melt-quenching technique proceeds through melting and casting routes for specific applications [68]. However, the major disadvantage associated with this technique is the problem associated with the presence of metallic ions forming undesirable compositions. The sol-gel technique is advantageous over melt-quenching technique in terms of expansion in terms of the compositional range at low processing temperatures with enhanced bioactivity of the system [22]. Moreover, doping of special ions with Bioglasses has been reported to enhance biological properties (mainly, antibacterial properties) [6], [27].

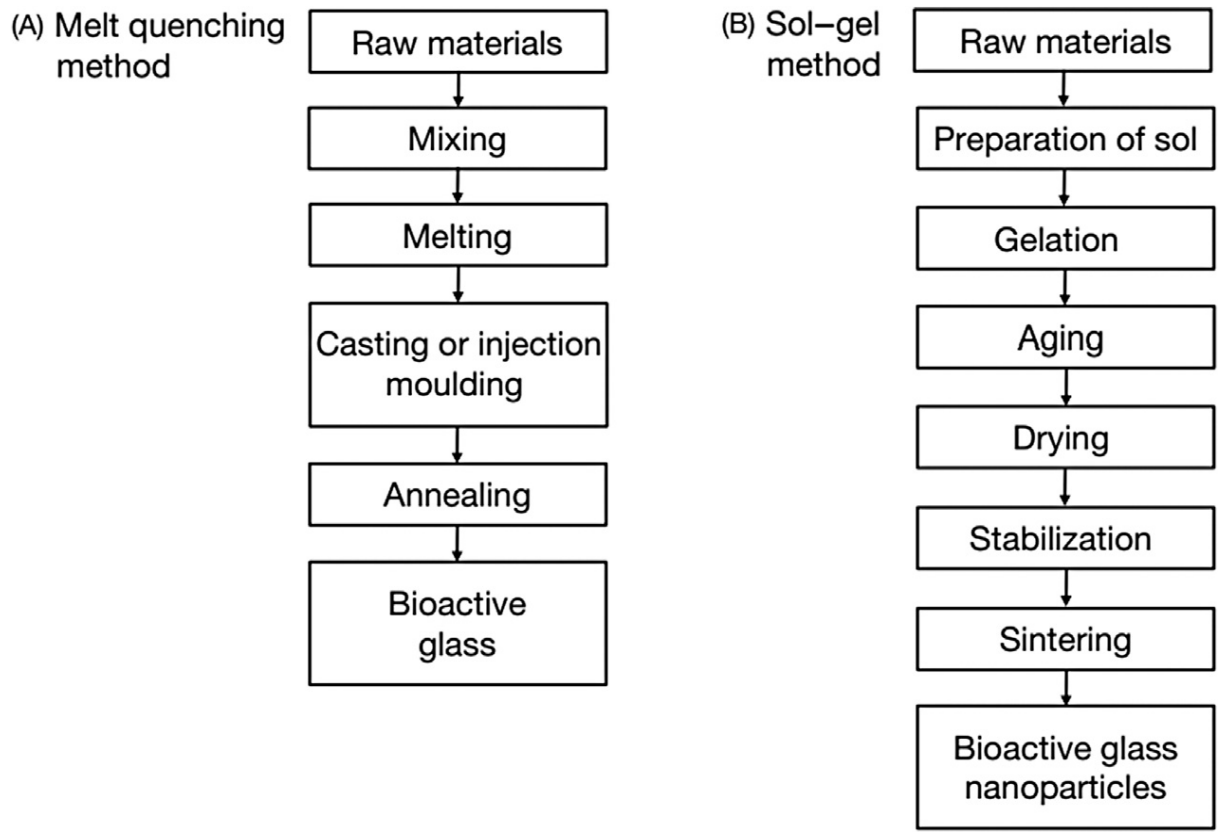

**Figure 9** Two main techniques for fabrication of bioactive glasses [6].

Nanoceramics have been reported to possess the highest tendency towards integration of cell and tissues owing to a very high surface area-to-volume ratio [6], [69]. For instance, nanoscale HA, has functional properties which are much enhanced as compared to those of microscale HA [6]. Moreover, nanoceramics possess high bioactivity owing to a high rate of dissolution of nanoscale structures, and a high surface area exposed to the biological environment (or, microenvironment, as such) as compared to that in microscale structures [70]. Besides, nanoceramics have been reported to possess excellent mechanical properties as compared to micro-sized ceramics at room temperature [71]. However, for load-bearing applications, the mechanical strength of HA ceramics is still found to be very low which restricts the application of these ceramics to metal implants where these are used as coatings on metal surface for the purpose of increasing implant biocompatibility and osteoconductivity [6], [71]–[73]. A reported technique to determine bioactivity of biomaterials is to immerse them in a simulated body fluid and investigate the formation of HA layer on the material surface after a certain time span at $37°C$ [72]–[74]. Zhang et al. [75] have investigated the influence of printing ink formulation of porous bioceramics during inkjet 3DP and reported that NP printing inks show a dramatic shrinkage in post-sintered condition, and hence, is not suitable for 3DP of bioceramics [75], [76]. Moreover, the printing ink formulation has also been reported to possess an influence on the mechanical properties of porous bioceramics. In addition, it has also been reported that with decrease in macroporosity, there is also a significant deterioration in terms of mechanical properties of these bioceramics [76].

5. **Piezoelectric nanoceramics**

A number of AM techniques such as Selective Laser Sintering (SLS), Three-Dimensional Printing (3DP), Stereolithography (SLA) and Laminated Object Manufacturing (LOM) have been used for fabrication of ceramic components with complex geometries [33]. The SLA technique is based on the principle of photopolymerization of liquid resins containing ceramic particles and has been extensively applied in fabrication of ceramic components owing to advantages offered in terms of high dimensional accuracy (or resolution) and precision in 3DP [77], [78]. For the purpose of describing the exact geometry of green parts and achieving dense components after photopolymerisation treatment, appropriate viscosity, solid loading and long-term stability (of the composite suspensions used in SLA process) are the three criterion to be satisfied. A high solid loading of ceramic powders ensures lower shrinkage and higher post-sintered density [79]. However, a high solid loading would adversely degrade the flowability of suspensions. Hence, a homogeneous distribution of ceramic filler materials in a given medium (for photopolymerisation) is necessary to ensure long-term stability and an enhancement of the fluidity of suspensions [80], [81].

In general, the presence of certain amounts of hydroxyl groups on the surface of ceramic particles makes them hydrophilic and subsequently highly prone to particle agglomeration [82]. Owing to the lower mechanical strength of green parts fabricated from aqueous suspensions, most SLA-based suspensions are hydrophobic systems [28]. A conversion of hydrophilic ceramic particle surfaces to hydrophobic surfaces through addition of polymeric surfactants has been reported through implementation of steric stabilisation mechanism [28]. One of the major challenges to be considered originates from the reduction of curing depth caused due to high solid loading of ceramic particles [83]. The ceramic particles in suspensions will lead to diminishing of the photopolymerisation reaction through dilution of the photosensitive medium and scattering of incident ultraviolet (UV) radiation [28]. The cure depth of photocurable suspensions may be described from relation (2) based on the Beer-Lambert Law [83]:

$$C_d = \left(\frac{2d_{50}}{3\phi}\right)\left(\frac{1}{Q}\right)\ln\left(\frac{E}{E_C}\right) \qquad (2)$$

where, $C_d$ is the depth of curing and is inversely proportional to the volume fraction of ceramic particles ϕ and the scattering coefficient Q and also directly proportional to the square of the refractive index (RI) difference ($\Delta n^2$) between the ceramic particle and photocurable liquid, E is the energy density associated with exposure in the resin, $E_c$ is the critical energy density or, in other words, the minimum input energy required to initiate photopolymerisation reaction [28]. Higher $\Delta n^2$ indicates a higher level of absorption and scattering of UV radiation by ceramic filler materials leading to a reduction in the depth of curing [28]. Besides, the particle size also largely influences the cure depth [28]. On one hand, nanoparticles have been reported to be favourable for attaining good piezoelectric properties owing to their high surface areas which can enhance the rate of sintering processes [28]. On the contrary, for a given volume, nanoceramic particles tend to enhance the scattering effect and thus, have been reported to decrease the cure depth due to a large number of scattering centres through lowering of particle size [84], [85].

Based on the work of Wang et al. [28], it has been shown that 3D printed ceramic specimens of $BaTiO_3$ exhibit grain sizes of the order of a few nanometres with a relative density of about 95% of the theoretical value [28]. Moreover, the fabricated nanoceramics with fine grain size show excellent dielectric properties and good piezoelectric constant (~ 163 pC/N as reported by Wang et al. [28]), which may be potentially exploited in ultra-capacitors and transducers with highly complex geometries [28].

## 6. Challenges and outlooks

### 6.1 From the industrial viewpoint

Preceramic resins such as siloxane resin system have been reported to possess an enormous potential in terms of replacing common acrylate resins for industrial purposes [22]. Although, pyrolysis of acrylate resins has previously been used to create glassy C, but the application of such glassy C is somewhat limited [24], [86]. The need of the hour is to design engineering systems which may be capable of fully exploiting the beneficial properties (especially mechanical properties) of TPP-SiOC, through the utilisation of its fine-grained microstructure [2]. The recently reported capability to fabricate centimetre-sized complex parts without sacrificing nanoscale accuracy shows potential for enhanced scalability in future [87], [88].

Moreover, the existence of a ductile-to-brittle transition for large-sized TPP-SiOC components remains to be investigated. This is because most of the application-oriented TPP-DLW-fabricated components consist of features with sizes at the micron scale and still have been reported to show an excellent combination of strength and ductility [39]. Besides, for a good plastic deformability in a complex-shaped part, it is highly essential to avoid surface-induced stress concentration which may otherwise encourage the material to fail in a brittle manner [2], [22], [24]. The fracture strength of a material is highly sensitive to localised stress concentration [87], [89]–[91]. Sharp notches at the nodes of nanolattices have been reported to limit their plastic deformability by leading to cracks at the nodes [92]. Designing a component with a good combination of strength and ductility remains a major challenge even today [93]–[100].

Although, as discussed in **Section 4**, nanoceramics possess enhanced bioactivity leading to a high level of tissue integration upon implantation, there exist a number of technical challenges in terms of their fabrication [6]. These primarily include high manufacturing cost, improper control of process parameters, and low yield of final products [74], [101]. This has paved the way towards the development of dual-phase nanoceramics for the purpose of overcoming the challenges associated with single-phase nanoceramics [27]. For instance, $Ca_3P$ with higher biodegradation capacity is mixed with nano-hydroxyapatite (nHA) in order to overcome degradation of nHA [6], [69]. Moreover, the properties of nanoceramics depend largely on the processing route and process parameters during fabrication. In addition, the main factors which influence the clinical efficiency of a biomaterial are its biocompatibility and functionality once implanted in the human body [71]. Thus, there exist a number of unexplored avenues (of multidisciplinary nature) in the field of nanoceramics and that an optimisation of process parameters is highly essential for a large-scale manufacture of high-quality nanoceramics at affordable costs. The term 'high-quality' is application-specific and covers a wide range of properties. Hence, from the viewpoint of industrial research, nanoceramics show an excellent potential towards interdisciplinary investigations in the near future.

## 6.2 From the fundamental viewpoint

Despite an enormous volume of published literature and reviews on 3DP of a number of materials, there is a limited information on the role of different kinds of microstructural defects on the structure-property correlation of materials, primarily due to lack of consistency in experimental investigations towards the direct visualization of these defects. This, of course, becomes highly interesting for polycrystalline nanoceramics. This is mainly due to the non-equilibrium cooling rates followed during 3DP techniques, thereby leading to a number of microstructural defects ranging from 0-D (point) defects (such as vacancies) to 3-

D (volume) defects (such as porosity, voids e.t.c.) [96], [102]–[104]. The most important of these are dislocations and microstructural boundaries which are 1-D (line) and 2-D (surface) defects, respectively [96]. The nano-scale interactions between dislocations and 2-D interfaces primarily control the mechanical response of the material [93]. Dislocations, in the context of crystalline materials, are associated with strain fields around them [88], [93], [105]–[108]. The nature of strain field is largely determined by the type of dislocations [109]–[112]. Hence, in the simplest terms, dislocations may be defined as regions of localized strain-induced curvature in a particular microstructure [113].

Based on this definition of dislocations, it may be further argued that there exist localised fluctuations in terms of crystallographic orientation of regions adjacent to a dislocation, from which the strain field around the same, maybe visualised using orientation and defect imaging microscopy techniques (such as Electron Backscatter Diffraction (EBSD)) for orientation imaging and either TEM or Electron Channeling Contrast Imaging (ECCI) for defect characterisation) coupled with theoretical strain tensor analysis [90]. It is worthwhile mentioning that TEM offers a much higher spatial resolution than ECCI, for characterisation of defects such as dislocations, stacking faults etc [90]. However, the primary limitations with defect imaging using TEM are complications associated with sample preparation and the field of imaging. In recent times, although the use of FIB-based sample preparation has enabled TEM to be used for imaging site-specific microstructural features, however, there is a huge amount of skill involved in preparation of TEM samples for site-specific studies [90]. In addition, the major problem involved in all aforementioned studies is the lack of reproducibility in experimental results. For instance, in a given microstructure, the arrangement of defects varies largely with the surface conditions. Hence, it is highly misleading to arrive at a conclusion on the defect arrangement based on a limited number 2D based investigations. In this context, it is always a scientifically good practice to perform a number of 3D investigations of microstructural and correlate the 2D information obtained with information in 3D for a proper understanding of microstructural defects. This, in a way, will also aid in unravelling the interaction between different defects during 3DP techniques. Understanding the 'defect-defect' interaction in materials (nanoceramics, in the present context) is highly essential for process parameter control during fabrication.

On the other hand, the simplest of microstructural boundaries are grain boundaries (GBs) which separate two grains (or, crystals) with difference in crystallographic orientation [114]. The structure of a GB is defined by five independent macroscopic degrees of freedom (DOFs) and three dependent microscopic DOFs [115]–[117]. This helps to classify different

GBs by defining a unique entity known as GB energy, in a polycrystalline material [115]–[117]. The local composition at GB also influences the mechanical response of the material. This gives rise to yet another entity known as 'GB excess' [115]–[117]. One of the main reasons as to why experimental investigations on the influence of defect interaction on the structure-property correlation of nanoceramics through "correlative microscopy" methodology is missing, is the complexity associated with the crystal structures of these materials and the enormous challenges associated with sample preparation for such investigations. This, in particular, offers an enormous potential wherein a number of future experimental cum theoretical investigations may be performed in order to understand structure-property correlation in nanoceramics based on atomic-scale analysis. In recent times, Artificial Intelligence (AI) and machine learning (ML) guided material design [115]–[117], are the avenues which offer an enormous potential in the tailoring of nanoceramic microstructures and aid material scientists to tailor a wide range of properties in these materials.

## 7. Conclusions

There is no doubt that in future, there will be a number of investigations (of multi-disciplinary nature) on exploring nanoceramics, ranging from regenerative medicine and tissue engineering to structural engineering. However, at present, this field is still pre-mature and before imagining any practical applications of nanoceramics, the primary step is to combine academic research involving a systematic understanding of structure-property correlation in these materials with industrial research based on overcoming the challenges involved in lowering of manufacturing cost and control of process parameters during fabrication processes to manufacture complex-shaped parts with mechanically toughened nanoceramics at an affordable cost and with high yield. Thus, an industry-academic collaboration is a must for progress in the field of nanoceramics.